\documentclass[aps,prb,10pt,twocolumn,superscriptaddress]{revtex4-2}
\usepackage[ascii]{inputenc}
\usepackage{amsmath,amssymb,amsfonts,amsthm}
\usepackage{mathtools}
\usepackage{tikz}
\usetikzlibrary{decorations.pathreplacing,decorations.pathmorphing}
\usepackage{braket}
\usepackage[caption=false]{subfig}
\usepackage{enumitem}
\usepackage[pdftex,bookmarks=false]{hyperref}

\DeclareMathOperator{\Tr}{Tr}
\DeclareMathOperator{\grad}{grad}
\DeclareMathOperator{\hess}{Hess}

\DeclareMathOperator{\asym}{skew}

\DeclareMathOperator*{\argmin}{argmin}

\DeclareMathOperator{\e}{e}

\newcommand{\N}{\mathbb{N}}
\newcommand{\Z}{\mathbb{Z}}
\newcommand{\R}{\mathbb{R}}
\newcommand{\C}{\mathbb{C}}
\newcommand{\A}{\mathbb{A}}

\DeclarePairedDelimiter\norm{\lVert}{\rVert}
\DeclarePairedDelimiter\inner{\langle}{\rangle}

\colorlet{gatecolor}{blue!15!white}

\def\gw{0.375}

\graphicspath{{figures/}}
\makeatletter
\def\input@path{{figures/}}
\makeatother

\begin{document}

\title{Riemannian quantum circuit optimization for Hamiltonian simulation}

\author{Ayse Kotil}
\email{ayse.kotil@tum.de}
\affiliation{Technical University of Munich, CIT, Department of Computer Science, Boltzmannstra{\ss}e 3, 85748 Garching, Germany}

\author{Rahul Banerjee}
\email{rahul.banerjee@tum.de}
\affiliation{Technical University of Munich, Department of Physics, James-Franck-Stra{\ss}e 1, 85748 Garching, Germany}
\affiliation{Technical University of Munich, CIT, Department of Computer Science, Boltzmannstra{\ss}e 3, 85748 Garching, Germany}

\author{Qunsheng Huang}
\email{keefe.huang@tum.de}
\affiliation{Technical University of Munich, CIT, Department of Computer Science, Boltzmannstra{\ss}e 3, 85748 Garching, Germany}

\author{Christian B.~Mendl}
\email{christian.mendl@tum.de}
\affiliation{Technical University of Munich, CIT, Department of Computer Science, Boltzmannstra{\ss}e 3, 85748 Garching, Germany}
\affiliation{Technical University of Munich, Institute for Advanced Study, Lichtenbergstra{\ss}e 2a, 85748 Garching, Germany}

\date{November 25, 2023}

\begin{abstract}
Hamiltonian simulation, i.e., simulating the real time evolution of a target quantum system, is a natural application of quantum computing. Trotter-Suzuki splitting methods can generate corresponding quantum circuits; however, a faithful approximation can lead to relatively deep circuits. Here we start from the insight that for translation invariant systems, the gates in such circuit topologies can be further optimized on classical computers to decrease the circuit depth and/or increase the accuracy. We employ tensor network techniques and devise a method based on the Riemannian trust-region algorithm on the unitary matrix manifold for this purpose. For the Ising and Heisenberg models on a one-dimensional lattice, we achieve orders of magnitude accuracy improvements compared to fourth-order splitting methods. The optimized circuits could also be of practical use for the time-evolving block decimation (TEBD) algorithm.
\end{abstract}

\maketitle

\section{Introduction}

Hamiltonian simulation is a natural and promising application of quantum computing \cite{Lloyd1996, Zalka1998}. For example, quantum time evolution gives access to the dynamical behavior of strongly correlated quantum systems, can quickly generate entanglement, and is a central ingredient of several quantum algorithms, like the HHL algorithm and quantum phase estimation.

A detailed numerical analysis of Trotter-Suzuki splitting methods \cite{Childs2021} shows that they can approximate the time evolution with circuit depth scaling essentially linearly in simulated time. Recently, the authors of \cite{Mansuroglu2023a, Tepaske2022, McKeever2022} have proposed and implemented the idea of optimizing variational circuit Ans\"atze inspired by Trotterized time evolution for the purpose of Hamiltonian simulation (using parametrized circuit gates). Here, we build upon the same idea and adapt a tensor network perspective as in \cite{McKeever2022}, but take a step further by regarding the circuit gates as general unitary matrices, analogous to \cite{Hauru2021, Geng2022}. Our main technical innovation is a derivation of how to employ the Riemannian trust-region algorithm \cite[chapter~7]{Absil2008} for the purpose of optimizing the quantum circuit.

Methods for approximating quantum time evolution on a quantum computer have also been explored in \cite{Low2017, Low2019, Haah2021, Barratt2020, Lin2021}. The authors of \cite{Barratt2020, Lin2021} focus on the best-approximation of the time-dependent quantum state, instead of the overall time evolution operator considered here. Approaches based on quantum signal processing \cite{Low2017, Low2019} might provide a complexity theoretic advantage in certain situations, but have the disadvantage of requiring additional auxiliary qubits and a block encoding of the Hamiltonian, which can incur a large overhead in practice.

\section{Notation and setup}

Consider the unitary time evolution operator (in units of $\hbar = 1$)
\begin{equation}
U(t) = \e^{-i H t}
\end{equation}
of a quantum system governed by a (time-independent) quantum Hamiltonian $H$ defined on a lattice. We denote the local dimension of each lattice site by $d$, i.e., the local Hilbert space is $\C^d$. In our numerical simulations we will set $d = 2$, but the method works for general $d$. Translation invariance of the Hamiltonian is assumed throughout.

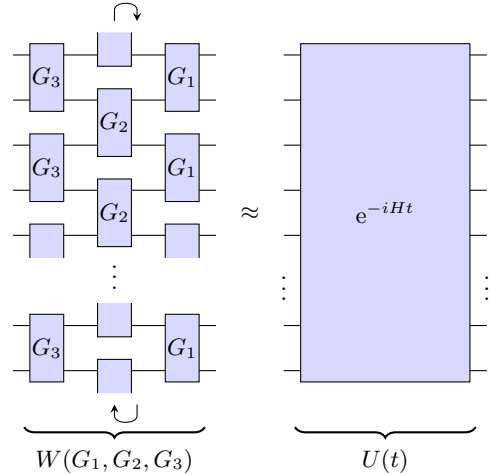
\begin{figure}[!ht]
\centering
\begin{tikzpicture}[scale=0.6, >=stealth]
\begin{scope}
\foreach \y in {0, 1, 3, 4, 5, 6, 7}
{
    \draw (-0.75, \y) -- ( 3.75, \y);
}
\foreach \x in {0, 2}
{
    \foreach \y in {0, 4, 6}
    {
        \draw[fill=gatecolor] ( 1.5*\x-\gw, \y-0.25) rectangle ( 1.5*\x+\gw, \y+1.25);
        \pgfmathsetmacro\z{int(3-\x)};
        \node at ( 1.5*\x, \y+0.5) {$G_\z$};
    }
    \draw[fill=gatecolor] ( 1.5*\x-\gw, 2.5) -- ( 1.5*\x-\gw, 3.25) -- ( 1.5*\x+\gw, 3.25) -- ( 1.5*\x+\gw, 2.5);
}
\foreach \x in {1}
{
    \foreach \y in {3, 5}
    {
        \draw[fill=gatecolor] ( 1.5*\x-\gw, \y-0.25) rectangle ( 1.5*\x+\gw, \y+1.25);
        \pgfmathsetmacro\z{int(\x+1)};
        \node at ( 1.5*\x, \y+0.5) {$G_\z$};
    }
    \node at ( 1.5*\x, 2.25) {$\vdots$};
    \draw[fill=gatecolor] ( 1.5*\x-\gw, 7.5) -- ( 1.5*\x-\gw, 6.75) -- ( 1.5*\x+\gw, 6.75) -- ( 1.5*\x+\gw, 7.5);
    \draw[fill=gatecolor] ( 1.5*\x-\gw, 1.5) -- ( 1.5*\x-\gw, 0.75) -- ( 1.5*\x+\gw, 0.75) -- ( 1.5*\x+\gw, 1.5);
    \draw[fill=gatecolor] ( 1.5*\x-\gw,-0.5) -- ( 1.5*\x-\gw, 0.25) -- ( 1.5*\x+\gw, 0.25) -- ( 1.5*\x+\gw,-0.5);
    \draw[->] (1.5*\x, 7.75) -- (1.5*\x, 8) to[out=90, in=90] (1.5*\x+0.5, 8) -- (1.5*\x+0.5, 7.75);
    \draw[->] (1.5*\x+0.5,-0.75) -- (1.5*\x+0.5,-1) to[out=-90, in=-90] (1.5*\x,-1) -- (1.5*\x,-0.75);
}
\draw[decorate, decoration={brace,mirror,amplitude=4}, thick] (-0.5,-1.25) -- ( 3.5,-1.25);
\node at ( 1.5, -2) {$W(G_1, G_2, G_3)$};
\end{scope}
\node at ( 4.5, 3.5) {$\approx$};
\begin{scope}[shift={( 6, 0)}]
\foreach \y in {0, 1, 3, 4, 5, 6, 7}
{
    \draw (-0.75, \y) -- ( 3.75, \y);
}
\draw[fill=gatecolor] (-\gw, -0.25) rectangle ( 3+\gw, 7.25);
\node at ( 1.5, 3.5) {$\e^{-i H t}$};
\node at (-0.75, 2) {$\vdots$};
\node at ( 3.75, 2) {$\vdots$};
\draw[decorate, decoration={brace,mirror,amplitude=4}, thick] (-0.5,-1.25) -- ( 3.5,-1.25);
\node at ( 1.5, -2) {$U(t)$};
\end{scope}
\end{tikzpicture}%
\caption{Example of a quantum circuit with brick wall layout and periodic boundary conditions, for approximating the exact time evolution operator.}
\label{fig:brickwall_circuit_approx}
\end{figure}

Our goal is to approximate $U$ by a quantum circuit. We designate the overall unitary transformation effected by the circuit as $W(G_1, \dots, G_n)$, with $G_\ell \in \mathcal{U}(d^2)$, $\ell = 1, \dots, n$, to-be optimized quantum gates forming the circuit, as illustrated in Fig.~\ref{fig:brickwall_circuit_approx} for a one-dimensional lattice and $n = 3$. Here and in the following, $\mathcal{U}(m)$ denotes the set of unitary $m \times m$ matrices. The circuit has a brick wall layout, and due to translation invariance, the gates within a layer are by construction all the same. We follow the mathematical convention of matrix chain ordering from right to left, i.e., the gates which are applied first are in the rightmost layer.

The Ansatz is motivated by the well-studied Trotterized time evolution approximation \cite{Childs2021}, which assumes that the Hamiltonian is a sum of ``simpler'' terms, $H = \sum_{\gamma=1}^{\Gamma} H_{\gamma}$, such that each $\e^{-i H_{\gamma} t}$ can be exactly realized as quantum circuit; a basic example is the even-odd splitting of a Hamiltonian with nearest-neighbor interactions on a one-dimensional lattice, $H = H_{\text{even}} + H_{\text{odd}}$, and the Strang splitting approximation
\begin{equation}
\label{eq:strang_splitting}
\e^{-i H t} = \e^{-i H_{\text{even}} t/2} \e^{-i H_{\text{odd}} t} \e^{-i H_{\text{even}} t/2} + \mathcal{O}(t^3).
\end{equation}
A benchmark comparison of our optimized circuits with such splitting methods is presented in Sect.~\ref{sec:simulations}.

We will devise a numerical method for optimizing the gates $G_\ell$ in sections \ref{sec:formalism_unitary_opt} and \ref{sec:method}. Each gate $G_\ell$ is regarded as general unitary matrix (instead of being parametrized), which opens up the broader range of Riemannian optimization techniques.

The time parameter $t$ is set to a fixed (model dependent) numerical value of order $1$. The optimized gates can then be used on a quantum computer to reach times which are integer multiples of $t$ by concatenating copies of the circuit. One expects that the approximation error increases only linearly with the final time \cite{Childs2021}.

\section{Light-cone considerations and generalization to larger systems}
\label{sec:light_cone}

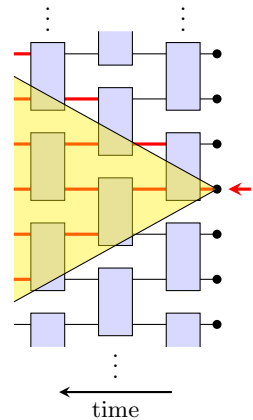
\begin{figure}[!ht]
\centering
\begin{tikzpicture}[scale=0.6, >=stealth, x=-1cm]
\begin{scope}
\foreach \y in {-3,...,3}
{
    \draw (-0.75, \y) -- ( 3.75, \y);
}
\draw[red, very thick] (-0.75,    0) -- ( 3.75, 0);
\draw[red, very thick] (     \gw, 1) -- ( 3.75, 1);
\draw[red, very thick] ( 1.5+\gw,-1) -- ( 3.75,-1);
\draw[red, very thick] ( 1.5+\gw, 2) -- ( 3.75, 2);
\draw[red, very thick] ( 3  +\gw,-2) -- ( 3.75,-2);
\draw[red, very thick] ( 3  +\gw, 3) -- ( 3.75, 3);
\draw[->, red, very thick] (-1.5, 0) -- (-1,    0);
\foreach \x in {0, 2}
{
    \foreach \y in {-2, 0, 2}
    {
        \draw[fill=gatecolor] ( 1.5*\x-\gw, \y-0.25) rectangle ( 1.5*\x+\gw, \y+1.25);
    }
    \node at ( 1.5*\x, 3.95) {$\vdots$};
    \draw[fill=gatecolor] ( 1.5*\x-\gw,-3.5) -- ( 1.5*\x-\gw,-2.75) -- ( 1.5*\x+\gw,-2.75) -- ( 1.5*\x+\gw,-3.5);
}
\foreach \x in {1}
{
    \foreach \y in {-3, -1, 1}
    {
        \draw[fill=gatecolor] ( 1.5*\x-\gw, \y-0.25) rectangle ( 1.5*\x+\gw, \y+1.25);
    }
    \node at ( 1.5*\x,-3.75) {$\vdots$};
    \draw[fill=gatecolor] ( 1.5*\x-\gw, 3.5) -- ( 1.5*\x-\gw, 2.75) -- ( 1.5*\x+\gw, 2.75) -- ( 1.5*\x+\gw, 3.5);
}
\end{scope}
\begin{scope}[shift={(-0.75, 0)}]
\foreach \y in {-3,...,3}
{
    \fill ( 0, \y) circle (0.1);
}
\draw[fill opacity=0.4, fill=yellow] ( 4.5,-2.5) -- ( 0, 0) -- ( 4.5, 2.5);
\draw[->, thick] ( 1,-4.5) -- node[below] {time} ( 3.5,-4.5);
\end{scope}
\end{tikzpicture}%
\caption{Physical light cone of causal correlations, and capability of a brick wall circuit to represent these (red bonds).}
\label{fig:brickwall_circuit_light_cone}
\end{figure}

For practical reasons, we will perform the circuit optimization for rather small system sizes; nevertheless, it turns out that the circuit is a faithful representation of the time evolution operator for larger systems as well (assuming translation invariance), as already noted in \cite{Heyl2019, Mansuroglu2023a}, cf.\ the detailed mathematical analysis in \cite{Haah2021}. To provide an intuitive argument why this works, we consider the light-cone picture shown in Fig.~\ref{fig:brickwall_circuit_light_cone}. The causal correlations spread with a finite velocity, and cannot exceed the Lieb-Robinson bounds \cite{LiebRobinson1972, ChenLucas2021}.

One observes that the following conditions have to be satisfied to arrive at a faithful approximation of the exact time evolution operator:
\begin{enumerate}[label=(\roman*)]
\item The evolution time $t$ has to be small enough such that the extent of the light cone at $t$ is smaller or equal to the system size. This property ensures that the light cone does not interfere with itself given the periodic boundary conditions.
\item The causal range of influence of the circuit gates (red thick lines in Fig.~\ref{fig:brickwall_circuit_light_cone}) has to enclose the physical light cone, to be able to represent the physical information spreading.
\end{enumerate}

Assuming these prerequisites hold, it becomes possible to use the circuit also for larger systems, simply by extending it with copies of the same gate $G_\ell$ in layer $\ell$. We will test this idea in our numerical experiments in Sect.~\ref{sec:simulations}. A practical use-case scenario consists of performing the optimization on a classical computer and small system size, and then using the optimized gates for larger systems on a quantum computer.

\section{Mathematical formalism of optimization on the manifold of unitary matrices}
\label{sec:formalism_unitary_opt}

In this section we briefly review the mathematical formalism for optimization under unitary constraints \cite{Edelman1998, Absil2008, Hauru2021}, following the notation in \cite{Absil2008}. This formalism forms the foundation for the numerical method proposed in the next section.

For fixed integer $m$, the set of unitary $m \times m$ matrices,
\begin{equation}
\mathcal{U}(m) = \{ V \in \C^{m \times m} \,\vert\, V^{\dagger} V = I_m \},
\end{equation}
with $I_m$ the $m \times m$ identity matrix, forms a mathematical manifold. $\mathcal{U}(m)$ can be interpreted as Riemannian submanifold of $\C^{m \times m}$ (with metric described below), and is a special case of the complex Stiefel manifold consisting of isometries. For notational simplicity, we omit the parameter $m$ from $\mathcal{U}(m)$ in the following.

A central concept is the \emph{tangent space} of a manifold. The tangent space at a given $V \in \mathcal{U}$ is parametrized by the set of complex anti-Hermitian matrices \cite{Absil2008}:
\begin{equation}
\label{eq:tangent_space_param}
T_V\mathcal{U} = \left\{ V A: A \in \C^{m \times m}, A^{\dagger} = -A \right\},
\end{equation}
where $A^{\dagger}$ denotes the adjoint (conjugate transpose) of $A$. By construction, $V^{\dagger} X$ is anti-Hermitian for any $X \in T_V\mathcal{U}$.

We introduce the following Riemannian metric on $T_V\mathcal{U}$, as in \cite{Hauru2021}:
\begin{equation}
\inner{\cdot, \cdot}_V: T_V\mathcal{U} \times T_V\mathcal{U} \to \R, \quad \inner{X, Y}_V = \Tr[X^{\dagger} Y].
\end{equation}
Note that $\Tr[X^{\dagger} Y] = \Tr[(V^\dagger X)^{\dagger} (V^\dagger Y)]$, and thus the trace is real-valued since $V^\dagger X$ and $V^\dagger Y$ are anti-Hermitian.

Viewing $\mathcal{U}$ as embedded into $\C^{m \times m}$, the corresponding projection onto the tangent space at $V \in \mathcal{U}$ reads \cite{Absil2008, Hauru2021}:
\begin{equation}
\label{eq:proj_tangent}
P_V X = V \asym(V^{\dagger} X),
\end{equation}
with $\asym(A) = \frac{1}{2} (A - A^{\dagger})$ the anti-Hermitian part of a matrix.

We define the \emph{gradient} of a smooth function $f: \C \to \R$ at point $z = x + i y$ with $x, y \in \R$ as composed of the derivatives with respect to the real and imaginary components of $z$:
\begin{equation}
\label{eq:def_grad}
\grad f(z) = \partial_x f(z) + i \partial_y f(z).
\end{equation}
This definition is straightforwardly generalized to functions depending on several complex numbers, e.g., $f: \C^m \to \R$ or $f: \C^{m \times m} \to \R$, by applying the definition entrywise.

We will encounter the situation that the to-be optimized target function $f: \mathcal{U} \to \R$ is the restriction of a function $\bar{f}$ defined on $\C^{m \times m}$. In this case, the gradient vector of $f$ results from projecting the gradient vector of $\bar{f}$ onto the tangent space:
\begin{equation}
\label{eq:grad_proj}
\grad f(V) = P_V \grad \bar{f}(V).
\end{equation}

For the purpose of computing second derivatives and Hessian matrices, we need some additional concepts. Let $\mathfrak{X}(\mathcal{U})$ denote the set of smooth vector fields on $\mathcal{U}$, following the notation of \cite{Absil2008}. We will use the unique \emph{Riemannian (Levi-Civita) connection} $\nabla$, which is formally defined as a map
\begin{equation}
\nabla: \mathfrak{X}(\mathcal{U}) \times \mathfrak{X}(\mathcal{U}) \to \mathfrak{X}(\mathcal{U}), \quad (\eta, \xi) \mapsto \nabla_{\eta} \xi
\end{equation}
which is symmetric and compatible with the Riemannian metric. Intuitively, $\nabla_{\eta}$ is the derivative of a vector field in direction $\eta$.

As before, we interpret $\mathcal{U}$ as Riemannian submanifold of $\C^{m \times m}$. Let $\xi \in \mathfrak{X}(\mathcal{U})$ be a vector field. Then the derivative of $\xi$ in gradient direction $X \in T_V \mathcal{U}$ ($V \in \mathcal{U}$) is given by \cite[Eq.~(5.15)]{Absil2008}
\begin{equation}
\label{eq:vector_field_derivative}
\nabla_X \xi = P_V (D\xi(V)[X]),
\end{equation}
where $D\xi(V)[X]$ is the gradient of $\xi$ in direction $X$ at point $V$.

A \emph{retraction} on $\mathcal{U}$ \cite[chapter~4]{Absil2008} is a mapping from the tangent bundle of the unitary matrix manifold into the manifold. We have found it convenient to use the polar decomposition ($V \in \mathcal{U}$) as retraction:
\begin{equation}
\label{eq:retraction_qr}
R: T\mathcal{U} \to \mathcal{U}, \quad R_V(\xi) = q_{\text{polar}}(V + \xi),
\end{equation}
where $q_{\text{polar}}(A)$ denotes the unitary matrix $Q \in \mathcal{U}$ from the polar decomposition of $A \in \C^{m \times m}$ as $A = Q P$, with $P$ a Hermitian positive semi-definite matrix of the same size as $A$. As a remark, alternative retraction methods have also been studied in the literature, based on QR-decompositions, projections, the Cayley transform and generally geodesic-like schemes \cite{Absil2012, Wen2013, Zhu2017, Geng2022}. We have found the polar decomposition to work well in practice, and leave a thorough exploration of alternative methods for future work.

In our numerical calculations, we will optimize the quantum gates $G_1, \dots, G_n$ simultaneously (instead of one after another). Matching this procedure with the mathematical formalism requires a generalization to target functions depending on several unitary matrices:
\begin{equation}
\label{eq:f_target_abstract}
f: \underbrace{\mathcal{U} \times \cdots \times \mathcal{U}}_{n \text{ terms}} \to \R.
\end{equation}
Formally, $f$ is a function from the product manifold $\mathcal{U}^{\times n}$ to the real numbers. The corresponding tangent space is the direct sum of the individual tangent spaces, cf.~\cite{Hauru2021}. In practice, the overall gradient vector is thus a concatenation of the individual gradient vectors in Eq.~\eqref{eq:grad_proj}, and the overall retraction results from applying the retraction in Eq.~\eqref{eq:retraction_qr} to the individual isometries and tangent vectors.

For minimizing $f$ in \eqref{eq:f_target_abstract}, we will use the Riemannian trust-region algorithm \cite[chapter~7]{Absil2008}. The central idea consists of a quadratic approximation of the target function in the neighborhood of a point $G \in \mathcal{U}^{\times n}$:
\begin{equation}
\hat{m}_G(X) = f(G) + \inner{\grad f(G), X} + \frac{1}{2} \inner{\hess f(G)[X], X}
\end{equation}
for $X \in T_G \mathcal{U}^{\times n}$, with the Riemannian Hessian
\begin{equation}
\label{eq:hessian_def}
\hess f(G)[X] = \nabla_X \grad f(G).
\end{equation}
The specific details for computing the gradient and Hessian will be discussed in the next section; with that, we have collected all ingredients for implementing the Riemannian trust-region algorithm \cite[Algorithm~10]{Absil2008}, using the truncated conjugate-gradient method for the trust-region subproblem, see \cite[Algorithm~11]{Absil2008} and \cite{Steihaug1983}.

\section{Numerical method for brick wall circuit optimization}
\label{sec:method}

As just mentioned, we will use the Riemannian trust-region algorithm \cite[chapter~7]{Absil2008} for the optimization. Here we describe the details for the specialization to the brick wall circuit Ansatz.

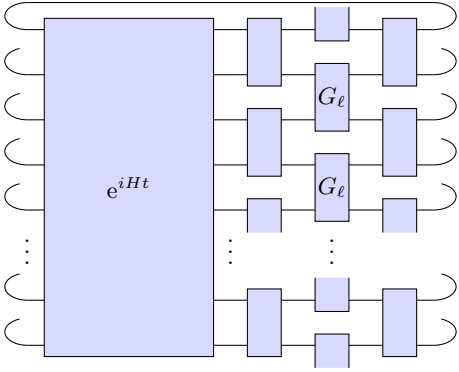
\begin{figure}[!ht]
\centering
\begin{tikzpicture}[scale=0.6, >=stealth]
\foreach \y in {0, 1, 3, 4, 5, 6}
{
    \draw (-0.75,       \y) -- ( 1.5*5+0.75, \y);
    \draw (-0.75,       \y) arc (270:110:0.5 and 0.3);
    \draw ( 1.5*5+0.75, \y) arc (-90:70:0.5 and 0.3);
}
\foreach \y in {7}
{
    \draw (-0.75,       \y    ) -- ( 1.5*5+0.75, \y    );
    \draw (-0.75,       \y+0.6) -- ( 1.5*5+0.75, \y+0.6);
    \draw (-0.75,       \y    ) arc (270:90:0.5 and 0.3);
    \draw ( 1.5*5+0.75, \y    ) arc (-90:90:0.5 and 0.3);
}
\begin{scope}[shift={( 1.5*3, 0)}]
\foreach \x in {0, 2}
{
    \foreach \y in {0, 4, 6}
    {
        \draw[fill=gatecolor] ( 1.5*\x-\gw, \y-0.25) rectangle ( 1.5*\x+\gw, \y+1.25);
    }
    \draw[fill=gatecolor] ( 1.5*\x-\gw, 2.5) -- ( 1.5*\x-\gw, 3.25) -- ( 1.5*\x+\gw, 3.25) -- ( 1.5*\x+\gw, 2.5);
}
\foreach \x in {1}
{
    \foreach \y in {3, 5}
    {
        \draw[fill=gatecolor] ( 1.5*\x-\gw, \y-0.25) rectangle ( 1.5*\x+\gw, \y+1.25);
        \pgfmathsetmacro\z{int(\x+1)};
        \node at ( 1.5*\x, \y+0.5) {$G_\ell$};
    }
    \node at ( 1.5*\x, 2.25) {$\vdots$};
    \draw[fill=gatecolor] ( 1.5*\x-\gw, 7.5) -- ( 1.5*\x-\gw, 6.75) -- ( 1.5*\x+\gw, 6.75) -- ( 1.5*\x+\gw, 7.5);
    \draw[fill=gatecolor] ( 1.5*\x-\gw, 1.5) -- ( 1.5*\x-\gw, 0.75) -- ( 1.5*\x+\gw, 0.75) -- ( 1.5*\x+\gw, 1.5);
    \draw[fill=gatecolor] ( 1.5*\x-\gw,-0.5) -- ( 1.5*\x-\gw, 0.25) -- ( 1.5*\x+\gw, 0.25) -- ( 1.5*\x+\gw,-0.5);
}
\end{scope}
\draw[fill=gatecolor] (-\gw, -0.25) rectangle ( 1.5*2+\gw, 7.25);
\node at ( 1.5,  3.5) {$\e^{i H t}$};
\node at (      -0.75, 2.25) {$\vdots$};
\node at ( 1.5*2+0.75, 2.25) {$\vdots$};
\end{tikzpicture}%
\caption{Tensor diagram representation of $\Tr[U^{\dagger} W(G)]$.}
\label{fig:brickwall_circuit_trace}
\end{figure}

We quantify the approximation error by the Frobenius norm distance $\norm{W - U}_{\mathrm{F}}^2$ between $U = \e^{-i H t}$ and the brick wall circuit $W$. (To shorten notation, we omit the explicit $t$ dependence of $U$.) We minimize this distance with respect to $G = (G_1, \dots, G_n)$, where $G_\ell \in \mathcal{U}(m)$ ($m = d^2$) for all $\ell = 1, \dots, n$:
\begin{equation}
G_{\text{opt}} = \argmin_{G \in \mathcal{U}(m)^{\times n}} \norm*{W(G) - U}_{\mathrm{F}}^2.
\end{equation}
Since both $U$ and $W(G)$ are unitary by construction, an expansion of the distance leads to
\begin{equation}
\label{eq:opt_distance}
\begin{split}
\norm{W - U}_{\mathrm{F}}^2
&= \Tr\big[(W - U)^{\dagger} (W - U)\big] \\
&= \Tr[I] - 2 \, \mathrm{Re}\Tr[U^{\dagger} W] + \Tr[I],
\end{split}
\end{equation}
where $I$ denotes the identity matrix. Thus we may equivalently minimize the following target function:
\begin{equation}
\label{eq:f_target}
f: \mathcal{U}(m)^{\times n} \to \R, \quad f(G) = -\mathrm{Re}\Tr[U^{\dagger} W(G)].
\end{equation}
A graphical tensor diagram representation of $\Tr[U^{\dagger} W(G)]$ is shown in Fig.~\ref{fig:brickwall_circuit_trace}.

The gradient of $f$ can be obtained as in Eq.~\eqref{eq:grad_proj}. The straightforward extension of $f$ reads
\begin{equation}
\bar{f}: \big(\C^{m \times m}\big)^{\times n} \to \R, \quad \bar{f}(G) = -\mathrm{Re}\Tr[U^{\dagger} W(G)],
\end{equation}
i.e., inserting general matrices into the brick wall diagram. In the present setting, since $f$ depends on several unitary matrices $G_1$, \dots, $G_n$, the corresponding individual projections $P_{G_\ell}$ ($\ell = 1, \dots, n$) have to be applied.

We make use of the Wirtinger formalism, summarized in appendix~\ref{sec:Wirtinger}, to obtain the gradient of $\bar{f}$. The Wirtinger derivative of $-\mathrm{Re}\Tr[U^{\dagger} W]$ with respect to an entry of $W$ is
\begin{multline}
\partial_{W_{jk}} (-1) \mathrm{Re}\Tr[U^{\dagger} W] \\
= \partial_{W_{jk}} (-1) \frac{1}{2} \left( \Tr[U^{\dagger} W] + \Tr[W^{\dagger} U] \right) %
= -\frac{1}{2} U_{jk}^*.
\end{multline}
Next, applying the chain rule \eqref{eq:chain_rule}, with $W$ regarded as function of $G_\ell$, leads to
\begin{equation}
\begin{split}
\partial_{G_\ell} \bar{f}(G) %
&= -\frac{1}{2} \sum_{j,k} U_{jk}^* \, \partial_{G_\ell} W_{jk}(G) \\
&= -\frac{1}{2} \Tr\!\left[U^{\dagger} \, \partial_{G_\ell} W(G)\right].
\end{split}
\end{equation}
Here we have used that $\partial_{G_\ell^*} W = 0$, such that the second term of the chain rule vanishes. The derivative of $W$ with respect to $G_\ell$ can be expressed as graphical diagram (shown for three layers) as
\begin{equation}
\begin{split}
\label{eq:brickwall_circuit_gradient}
&\partial_{G_\ell} W = \\
&\begin{tikzpicture}[scale=0.6]
\useasboundingbox (-0.75,-1.5) rectangle (11.5, 5.5);
\begin{scope}
\foreach \y in {0,...,4}
{
    \draw (-0.75, \y) -- ( 3.75, \y);
}
\foreach \x in {0, 2}
{
    \foreach \y in {1, 3}
    {
        \draw[fill=gatecolor] ( 1.5*\x-\gw, \y-0.25) rectangle ( 1.5*\x+\gw, \y+1.25);
    }
    \draw[fill=gatecolor] ( 1.5*\x-\gw,-0.5) -- ( 1.5*\x-\gw, 0.25) -- ( 1.5*\x+\gw, 0.25) -- ( 1.5*\x+\gw,-0.5);
}
\draw[fill=gatecolor] ( 1.5-\gw, 4.5) -- ( 1.5-\gw, 3.75) -- ( 1.5+\gw, 3.75) -- ( 1.5+\gw, 4.5);
\draw[dashed, fill=white] ( 1.5-\gw, 2-0.25) rectangle ( 1.5+\gw, 3.25);
\draw[fill=gatecolor] ( 1.5-\gw, -0.25) rectangle ( 1.5+\gw, 1.25);
\node at ( 1.5, 0.5) {$G_\ell$};
\node at ( 1.5,-1)    {$\vdots$};
\node at ( 1.5, 5.25) {$\vdots$};
\end{scope}
\node at ( 4.5, 2) {$+$};
\begin{scope}[shift={( 6, 0)}]
\foreach \y in {0,...,4}
{
    \draw (-0.75, \y) -- ( 3.75, \y);
}
\foreach \x in {0, 2}
{
    \foreach \y in {1, 3}
    {
        \draw[fill=gatecolor] ( 1.5*\x-\gw, \y-0.25) rectangle ( 1.5*\x+\gw, \y+1.25);
    }
    \draw[fill=gatecolor] ( 1.5*\x-\gw,-0.5) -- ( 1.5*\x-\gw, 0.25) -- ( 1.5*\x+\gw, 0.25) -- ( 1.5*\x+\gw,-0.5);
}
\draw[fill=gatecolor] ( 1.5-\gw, 4.5) -- ( 1.5-\gw, 3.75) -- ( 1.5+\gw, 3.75) -- ( 1.5+\gw, 4.5);
\draw[fill=gatecolor] ( 1.5-\gw, 1.75) rectangle ( 1.5+\gw, 3.25);
\node at ( 1.5, 2.5) {$G_\ell$};
\draw[dashed, fill=white] ( 1.5-\gw,-0.25) rectangle ( 1.5+\gw, 1.25);
\node at ( 1.5,-1)    {$\vdots$};
\node at ( 1.5, 5.25) {$\vdots$};
\end{scope}
\node at ( 10.5, 2) {$+$};
\node at ( 11.5, 2) {$\cdots$};
\end{tikzpicture}

\end{split}
\end{equation}
The uncontracted legs of the dashed ``holes'' in the network form the gradient tensor. The summation is due to the product rule. The tensor network diagram of $\Tr[U^{\dagger} \, \partial_{G_\ell} W(G)]$ then results from combining Fig.~\ref{fig:brickwall_circuit_trace} with Eq.~\eqref{eq:brickwall_circuit_gradient}. Note that $\Tr[U^{\dagger} \, \partial_{G_\ell} W(G)]$ has the same dimensions as $G_\ell$ (instead of a scalar quantity, as the trace might suggest).

Finally, we can use relation \eqref{eq:gradient_Wirtinger_relation} to obtain the gradient of $\bar{f}$ with respect to the unitary matrices $(G_1, \dots, G_n)$:
\begin{equation}
\grad \bar{f}(G) = - \Tr\!\left[U^{\dagger} \, \partial_{G_\ell} W(G)\right]^*_{\ell = 1, \dots, n}.
\end{equation}
The gradient of $f$ is then, according to Eq.~\eqref{eq:grad_proj},
\begin{equation}
\label{eq:grad_f}
\grad f(G) = - P_{G_\ell} \Tr\!\left[U^{\dagger} \, \partial_{G_\ell} W(G)\right]^*_{\ell = 1, \dots, n}.
\end{equation}

In practice, we have found it convenient to work with a real-valued representation of the gradient. For that purpose, we first parametrize the tangent spaces $T_{G_\ell} \mathcal{U}$, $\ell = 1, \dots, n$: let
\begin{equation}
\A_m = \left\{ A \in \C^{m \times m} \vert A^{\dagger} = -A \right\}
\end{equation}
denote the set of anti-Hermitian $m \times m$ matrices (which is a vector space over the real numbers). Then the following map defines an isometry between this space and the real-valued $m \times m$ matrices:
\begin{equation}
\mathfrak{s}: \R^{m \times m} \to \A_m, \quad \mathfrak{s}(R) = \frac{1}{2} \left(R - R^T\right) + \frac{i}{2} \left(R + R^T\right),
\end{equation}
with inverse $\mathfrak{s}^{-1}(A) = \mathrm{Re}(A) + \mathrm{Im}(A)$. $\mathfrak{s}$ preserves the inner product $\inner{A, B} = \Tr[A^{\dagger} B]$. Together with Eqs.~\eqref{eq:tangent_space_param} and \eqref{eq:proj_tangent}, we may thus isometrically map the gradient of $f$ to a list of real-valued matrices:
\begin{multline}
\label{eq:grad_f_real}
\widetilde{\grad} f(G) = \\
- \mathfrak{s}^{-1}\left(\asym\left({G_\ell}^{\dagger} \Tr\!\left[U^{\dagger} \, \partial_{G_\ell} W(G)\right]^*\right)\right)_{\ell = 1, \dots, n}.
\end{multline}
For convenience, we finally reshape this gradient into a real vector of length $n m^2$ in our calculations.

For calculating the Hessian appearing in Eq.~\eqref{eq:hessian_def}, note that $X \in T_G \mathcal{U}^{\times n}$ consists of a list of tangent vectors: $X = (X_1, \dots, X_n)$ with $X_\ell \in T_{G_\ell}\mathcal{U}$ for $\ell = 1, \dots, n$. Combined with the formula \eqref{eq:grad_f}, we have to evaluate
\begin{equation}
\label{eq:hessian_terms}
- \nabla_{X_\ell} P_{G_{\ell'}} \Tr\!\left[U^{\dagger} \, \partial_{G_{\ell'}} W(G)\right]^*
\end{equation}
for all $\ell, \ell' = 1, \dots, n$. We use Eq.~\eqref{eq:vector_field_derivative} for that purpose. In case $\ell \neq \ell'$, the derivative in direction $X_\ell$ takes a similar form as in Eq.~\eqref{eq:brickwall_circuit_gradient}, but with the ``holes'' in layer $\ell$ filled by $X_\ell$. In case $\ell = \ell'$, there is a contribution from the trace in Eq.~\eqref{eq:hessian_terms}, formed by replacing one of the remaining $G_\ell$ matrices by $X_\ell$ and summing over all occurrences. Another contribution stems from the projector $P_{G_\ell}$: for this, we first rewrite the definition in Eq.~\eqref{eq:proj_tangent} as
\begin{equation}
P_{G_\ell} Z = \frac{1}{2} Z - \frac{1}{2} G_{\ell} Z^{\dagger} G_{\ell}, \quad Z \in \C^{m \times m}.
\end{equation}
Thus the gradient of $P_{G_\ell} Z$ in direction $X_\ell$ ($Z$ regarded constant) equals
\begin{equation}
D(P_{G_\ell} Z)[X_\ell] = -\frac{1}{2} \left( X_{\ell} Z^{\dagger} G_{\ell} + G_{\ell} Z^{\dagger} X_{\ell} \right).
\end{equation}
Analogous to the gradient in Eq.~\eqref{eq:grad_f_real}, we have found it convenient to use the map $\mathfrak{s}$ to parametrize the tangent vectors $X_\ell$ in terms of real $m \times m$ matrices, and accordingly represent the Hessian as real symmetric $n m^2 \times n m^2$ matrix. Note that the Hessian matrix is (in general) not positive semidefinite.

As mentioned above, based on the gradient in Eq.~\eqref{eq:grad_f_real} and the real-valued Hessian matrix, we now employ the Riemannian trust-region algorithm \cite[Algorithm~10]{Absil2008} combined with the truncated conjugate-gradient method for the trust-region subproblem \cite[Algorithm~11]{Absil2008}, \cite{Steihaug1983} to minimize the target function \eqref{eq:f_target} with respect to the unitary matrices $G = (G_1, \dots, G_n) \in \mathcal{U}(m)^{\times n}$. The hyperparameters of the algorithm are chosen as: initial radius $\Delta_0 = 0.01$, maximum radius $\bar{\Delta} = 0.1$, and $\rho'= \frac{1}{8}$.

The optimization is sensitive to the initial brick wall unitaries used as starting point, and is not always converging to the global optimum according to our numerical experiments. We have found the following two strategies useful for obtaining expedient starting values: (i) A ``bootstrapping'' approach, using the optimized gates from a circuit with fewer layers (typically two less) and padding identity layers (i.e., containing identity matrices) on the left and right. All the gates are then optimized simultaneously. (ii) Employing a splitting method from the literature as starting point. Given a Hamiltonian with two-body interactions, a splitting method provides two-qubit gates which form the same brick wall topology as our optimization Ansatz. By using these gates as starting point, the optimized circuit performs at least as good as the splitting method. For comparison, we demonstrate the effect of a more simplistic starting point, namely all identity matrices, for the Heisenberg model (see below).

The Python/NumPy source code of our implementation, including the numerical experiments of the following section, is available at \cite{rqcopt}. We have tested the gradient and Hessian computation by comparison with finite difference approximations of derivatives.

\section{Numerical simulations}
\label{sec:simulations}

As demonstration, we apply the numerical method in section~\ref{sec:method} to several quantum lattice models. To describe the physical setup, first consider a Hamiltonian on the one-dimensional lattice $\Z_{/(L)}$ with $L$ sites (enumerated as $0, 1, \dots, L-1$) and periodic boundary conditions:
\begin{equation}
H = \sum_{j=0}^{L-1} \hat{h}_{j,j+1}.
\end{equation}
Here the subscripts indicate that the operator $\hat{h}$ acts locally on sites $j, j+1$.

A natural benchmark is a comparison with Trotterized even-odd splitting schemes. On a one-dimensional lattice, the Hamiltonian is partitioned into an even and odd part, namely $H = H_{\text{even}} + H_{\text{odd}}$ with
\begin{equation}
H_{\text{even}/\text{odd}} = \sum_{\substack{j=0 \\ j \text{ even/odd}}}^{L-1} \hat{h}_{j,j+1}.
\end{equation}
By construction, the summands commute pairwise since the operators act on disjoint sites. A time step $\Delta t$ of $H_{\text{even}/\text{odd}}$ is exactly realized by a quantum circuit layer consisting of copies of the two-qubit quantum gate $\e^{-i \hat{h} \Delta t}$. A step of the Strang splitting approximation, Eq.~\eqref{eq:strang_splitting}, can then be implemented using three circuit layers with the same layout as in Fig.~\ref{fig:brickwall_circuit_approx}. The construction works analogously for other splitting methods.

We fix a time $t$ and quantify the approximation error by the spectral norm distance between the numerically exact time evolution operator $\e^{-i H t}$ and the unitary matrix resulting from the splitting method or optimized circuit $W(G_1, \dots, G_n)$, respectively. Note that the target function \eqref{eq:f_target} actually minimizes the Frobenius norm distance, since this has a canonical and simpler tensor network representation, as shown in Fig.~\ref{fig:brickwall_circuit_trace}.

For existing splitting schemes we consider the approximation error for an increasing number of steps $r \in \N_{\ge 1}$. Thus a single time step has size $\Delta t = \frac{t}{r}$ and consists, e.g., of three substeps for the Strang method. We choose the convention that the ``cost'' of an integration method is the number $n$ of required circuit layers. Since we can always merge the last substep with the first substep of the next time step, the number of layers for the Strang method is $n = 2 r + 1$. In general, for a method with $s$ substeps, we have $n = (s - 1) r + 1$.

Besides the Strang method, we also apply the Suzuki formula of order $4$ by M.~Suzuki \cite{Suzuki1991}, the method of order $4$ by H.~Yoshida \cite{Yoshida1990}, the Runge-Kutta-Nystr\"om method of order $4$ by R.~I.~McLachlan \cite{McLachlan1995}, and the partitioned Runge-Kutta method $\text{S}_6$ of order $4$ by S.~Blanes and P.~C.~Moan \cite{Blanes2002}.

\subsection{Ising model on a one-dimensional lattice}

We first consider the transverse-field Ising model (TFIM) Hamiltonian on the one-dimensional lattice $\Z_{/(L)}$ with $L$ sites (enumerated as $0, 1, \dots, L-1$) and periodic boundary conditions:
\begin{equation}
\label{eq:H_Ising}
H^{\text{Ising}} = \sum_{j=0}^{L-1} \left( J Z_j Z_{j+1} + g X_j + h Z_j \right).
\end{equation}
Here $X_j$ and $Z_j$ are the usual Pauli matrices acting on site $j$, and $J, g, h \in \R$ are parameters. Without loss of generality, we set $J = 1$. We will first consider the \emph{integrable} case $h = 0$, and then investigate general parameter values for which the model becomes non-integrable.

\begin{figure}[!ht]
\centering
\subfloat[$5$ circuit layers]{%
\label{fig:ising1d_dynamics_opt_visualization_n5}%
\includegraphics[width=\columnwidth]{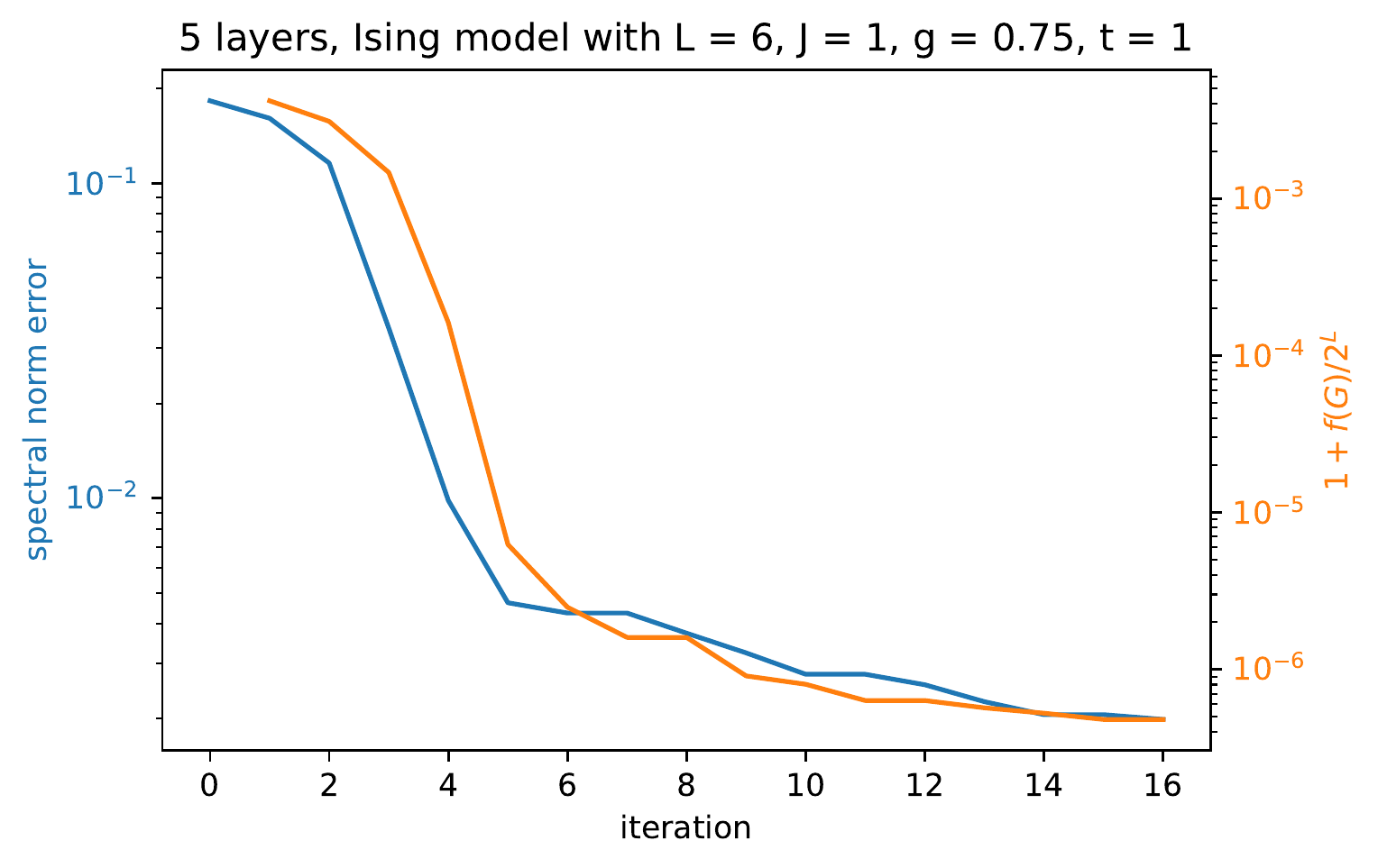}} \\
\subfloat[$7$ circuit layers]{%
\label{fig:ising1d_dynamics_opt_visualization_n7}%
\includegraphics[width=\columnwidth]{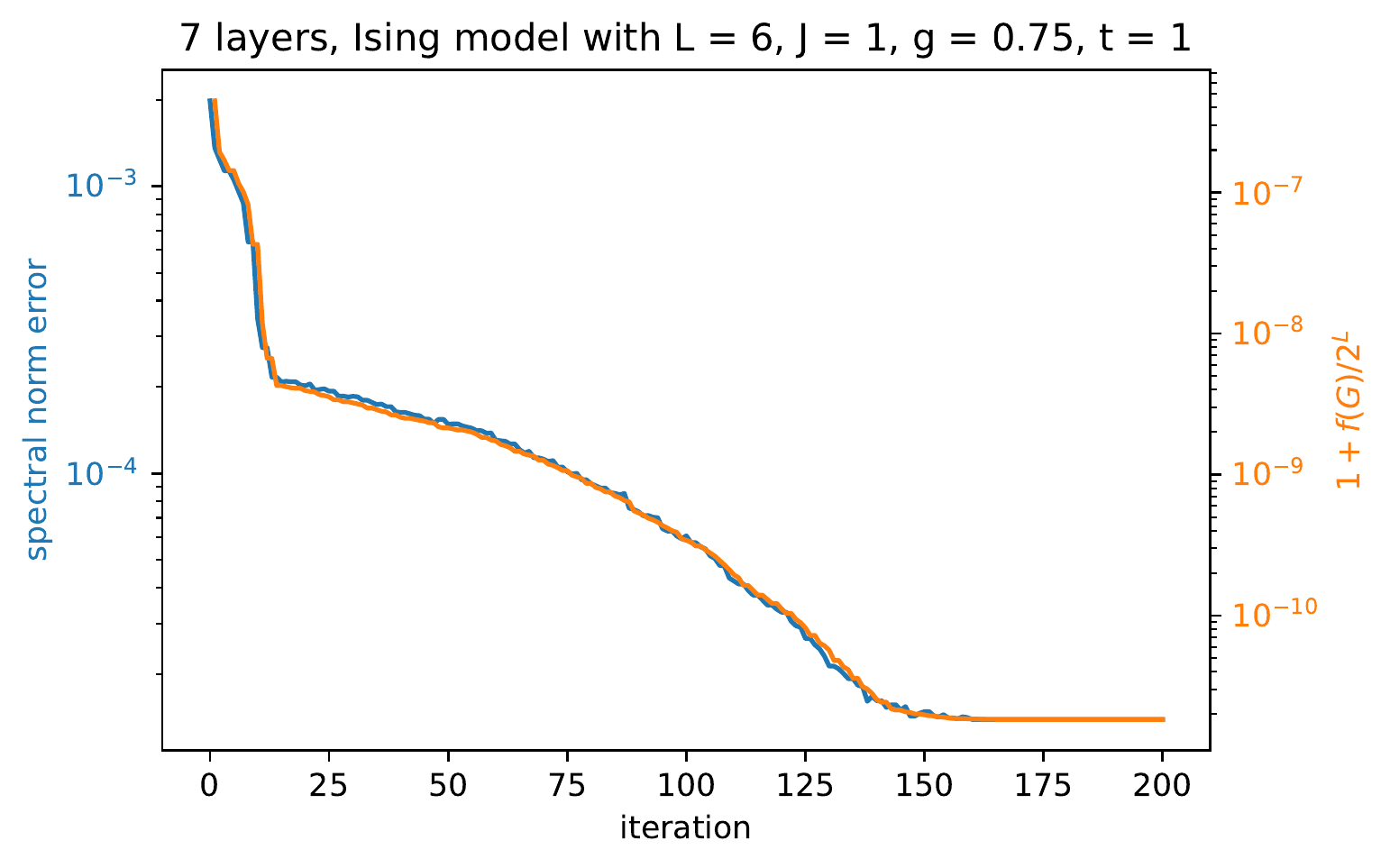}}
\caption{Progress of the Riemannian trust-region algorithm iteration, for the Ising model Hamiltonian \eqref{eq:H_Ising} on a one-dimensional lattice with $L = 6$ sites, $h = 0$ and time $t = 1$. The blue curves shows the approximation error quantified by the spectral norm distance, and the orange curves the shifted and rescaled target function $f$ in Eq.~\eqref{eq:f_target}.}
\label{fig:ising1d_dynamics_opt_visualization}
\end{figure}

\begin{figure}[!htp]
\centering
\subfloat[integrable case ($h = 0$)]{%
\label{fig:ising1d_ordered_dynamics_circuit_error}%
\includegraphics[width=\columnwidth]{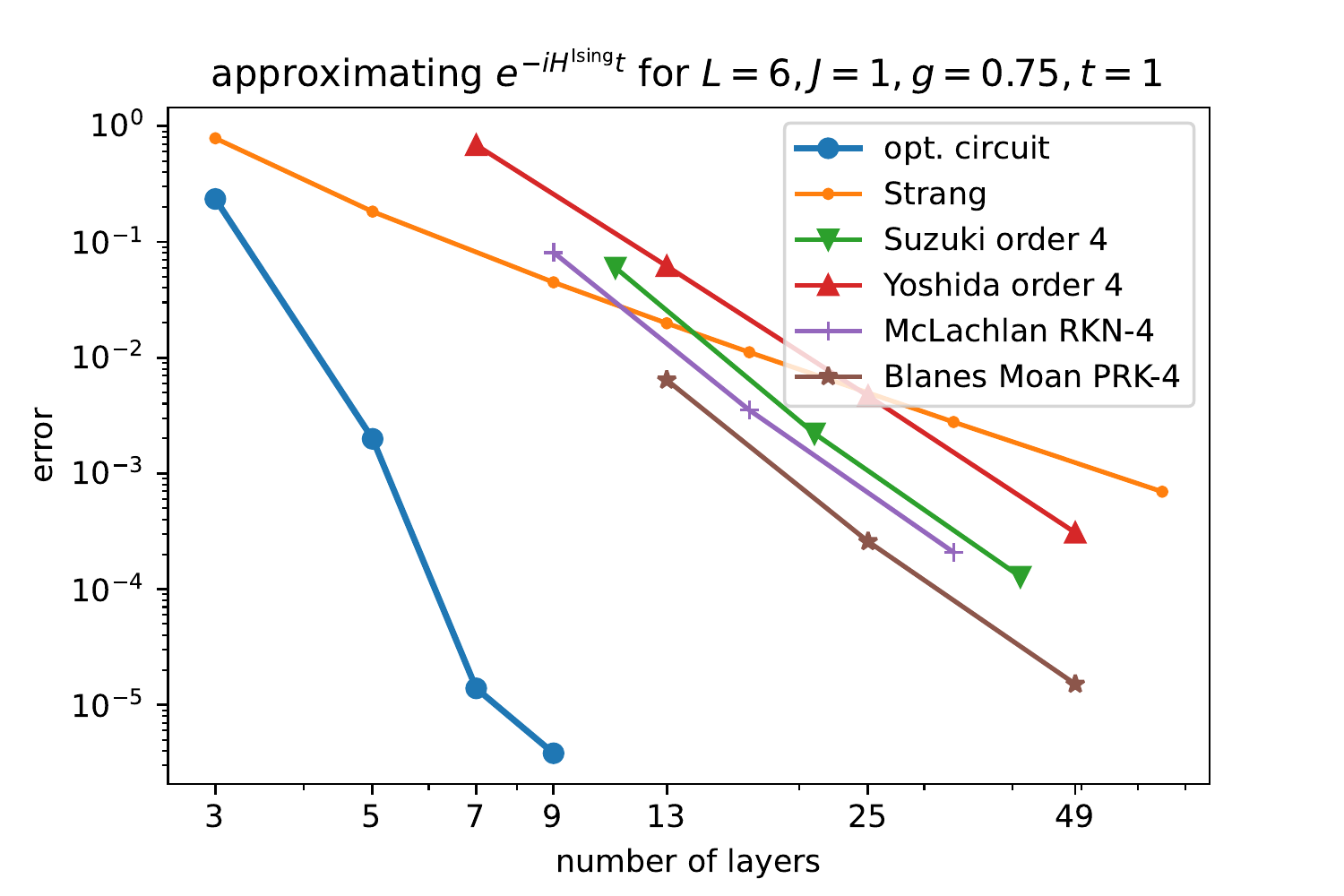}} \\
\subfloat[dependence on system size $L$]{%
\label{fig:ising1d_dynamics_approx_larger_systems}%
\includegraphics[width=\columnwidth]{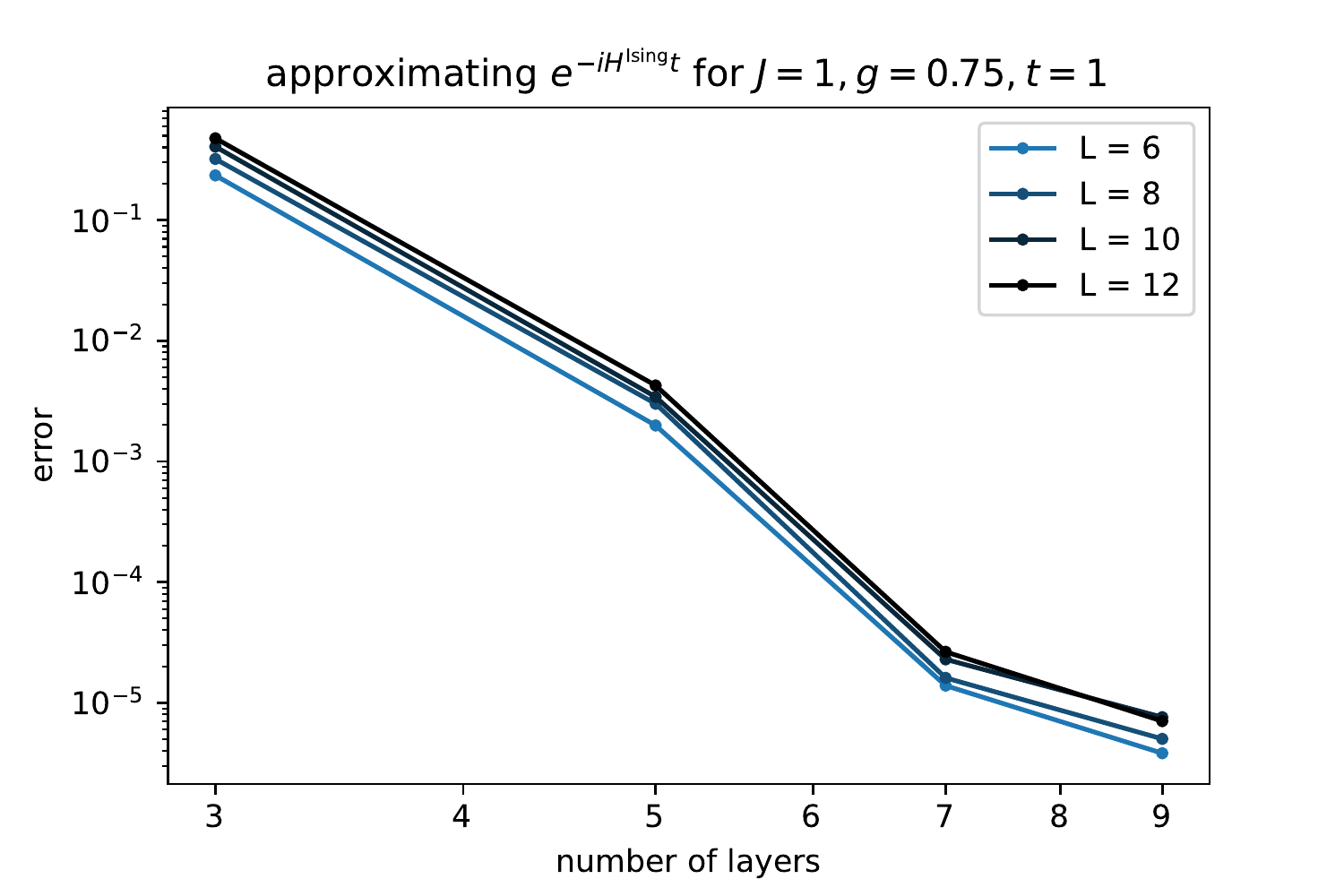}} \\
\subfloat[non-integrable case ($h = 0.6$)]{%
\label{fig:ising1d_nonintegrable_dynamics_circuit_error}%
\includegraphics[width=\columnwidth]{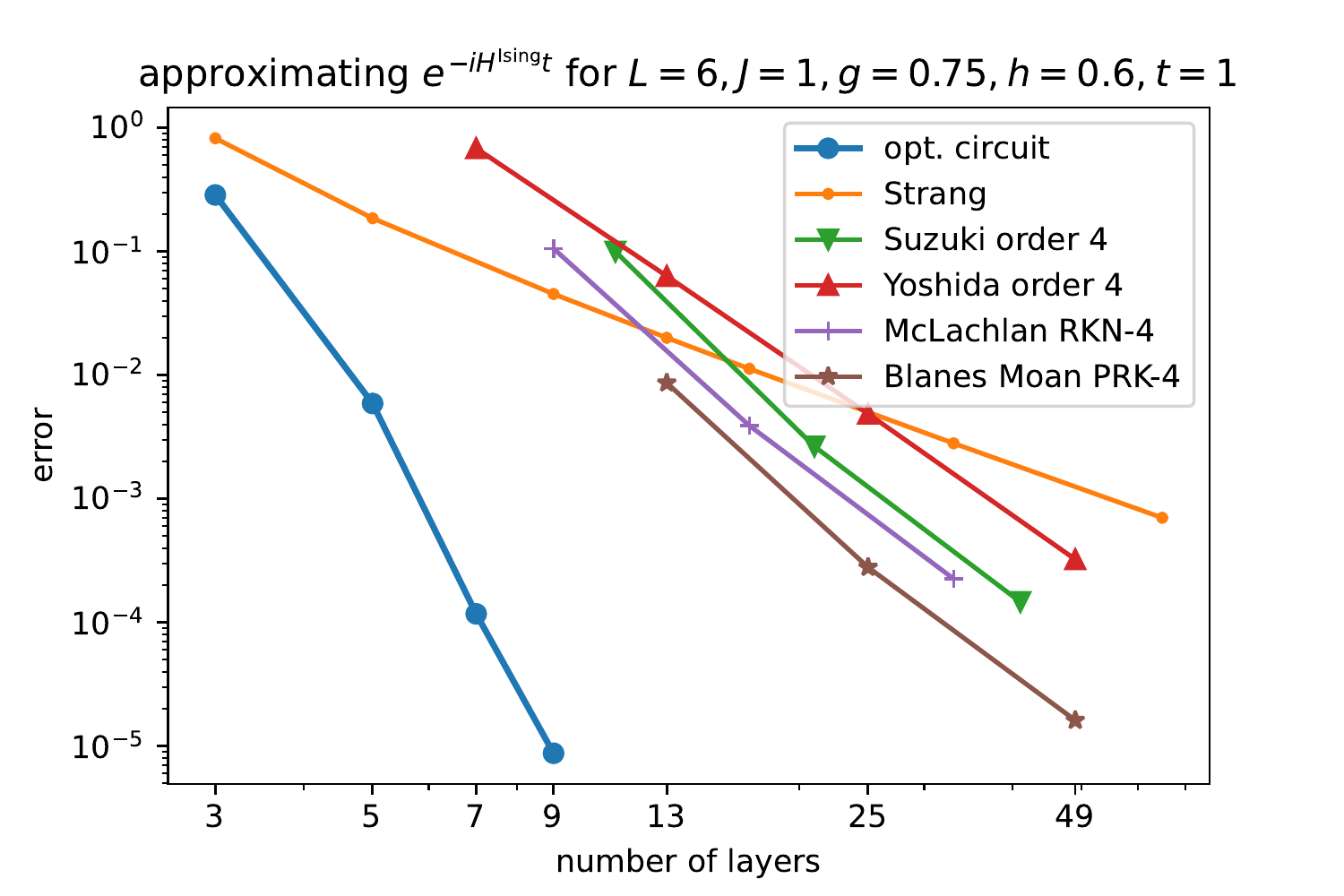}}
\caption{Approximation error (quantified by the spectral norm distance) of the quantum time evolution operator for $t = 1$ governed by the Ising model Hamiltonian \eqref{eq:H_Ising} on a one-dimensional lattice. The circuit gates have been optimized for $L = 6$ lattice sites. (a) and (c) show a comparison with existing splitting methods from the literature, with the thick blue curves corresponding to the optimized circuits of the present work. Subplot (b) displays the error for larger systems, always using the same optimized brickwall circuit gates $(G_1, \dots, G_n)$ found in (a).}
\label{fig:ising1d_dynamics_circuit_error}
\end{figure}

To demonstrate the behavior of the Riemannian trust-region algorithm, Fig.~\ref{fig:ising1d_dynamics_opt_visualization} visualizes the spectral norm distance and target function during the optimization iterations. For $7$ circuit layers, the two curves almost overlap exactly. One also notices that a plateau is reached after around 160 iterations, such that we stop after 200 iterations. We have adopted the strategy of either using the circuit gates of an existing splitting method as starting point for the optimization, or start with the optimized gates from a circuit with less layers and pad identity gates (which are then subject to the optimization as well).

A benchmark evaluation of the numerical optimization for the integrable Ising model and $t = 1$ is shown in Fig.~\ref{fig:ising1d_dynamics_circuit_error}. For the even-odd splitting methods, we have used the two-qubit gate $\e^{-i J \left( Z \otimes Z + g \frac{1}{2}(X \otimes I_2 + I_2 \otimes X) \right) \Delta t}$. The approximation error of the optimized circuit is represented by the thick blue curve. One observes a clear advantage: for example, the optimized circuit achieves the same or better accuracy using $9$ layers as compared to the $\text{S}_6$ method by S.~Blanes and P.~C.~Moan using $49$ layers. The chosen parameter $g = 0.75$ corresponds to the ordered phase; as indication that the results are not parameter-specific, we have repeated the calculations for $g = 1.5$ corresponding to the disordered phase, with qualitatively same outcome (data not shown).

As discussed at the end of Sect.~\ref{sec:light_cone}, it is possible to use the optimized circuit gates also for larger system sizes $L$. The corresponding approximation error quantified by the spectral norm distance is shown in Fig.~\ref{fig:ising1d_dynamics_approx_larger_systems}. As expected from the light cone picture, the error hardly increases for larger $L$.

Naturally, one could attribute the large effectiveness of the circuit optimization to the integrable property of the TFIM. In this sense, our method could be regarded as numerical equivalent of circuit compression based on the integrable structure manifest in the Yang-Baxter equation, as studied in Refs.~\cite{Peng2022, Kokcu2022, Camps2022, Astrakhantsev2023}. Surprisingly, however, the optimization results remain qualitatively unaffected by turning on an integrability-breaking longitudinal field, i.e., setting $h$ to a non-zero value, as shown in Fig.~\ref{fig:ising1d_nonintegrable_dynamics_circuit_error}.

\subsection{Heisenberg model on a one-dimensional lattice}

As next example, we consider the Heisenberg-type Hamiltonian
\begin{equation}
\label{eq:H_Heisenberg}
H^{\text{Heis}} = \sum_{j=0}^{L-1} \sum_{\alpha = 1, 2, 3} \left( J_\alpha \sigma^{\alpha}_j \sigma^{\alpha}_{j+1} + h_\alpha \sigma^{\alpha}_j \right),
\end{equation}
with $(\sigma^1, \sigma^2, \sigma^3) = (X, Y, Z)$ the vector of Pauli matrices and $\vec{J} \in \R^3$, $\vec{h} \in \R^3$ parameters. In our numerical simulations we set $\vec{J} = (1, 1, -\frac{1}{2})$ and $\vec{h} = (\frac{3}{4}, 0, 0)$. As before, the model is defined on a lattice with $L$ sites and periodic boundary conditions. The final time is set to $t = \frac{1}{4}$, which is smaller than for the Ising model to compensate for the faster spreading velocity of correlations (data not shown).

\begin{figure}[!ht]
\centering
\subfloat[comparison with splitting methods]{%
\label{fig:heisenberg1d_dynamics_circuit_error}%
\includegraphics[width=\columnwidth]{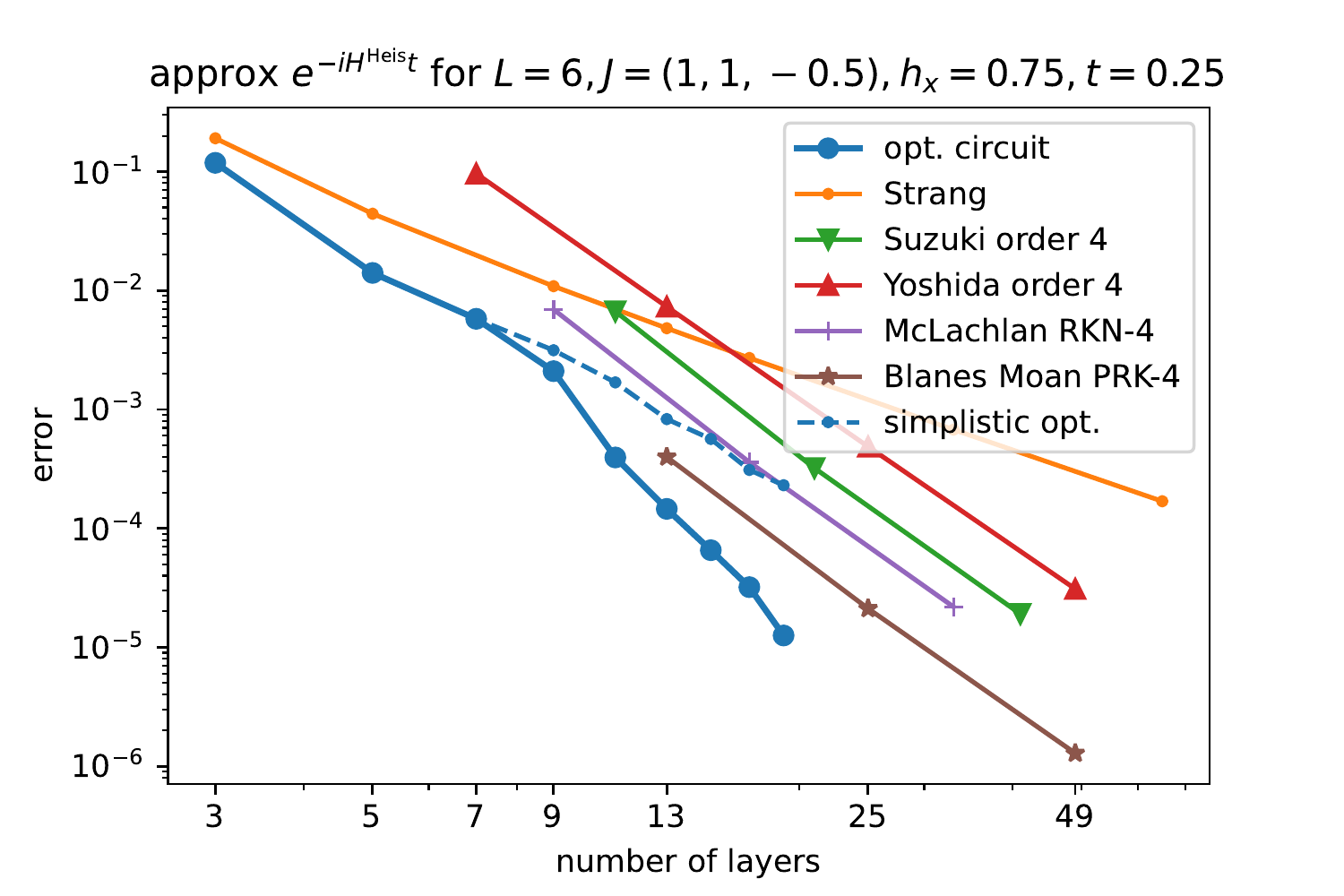}}\\
\subfloat[dependence on system size $L$]{%
\label{fig:heisenberg1d_dynamics_approx_larger_systems}%
\includegraphics[width=\columnwidth]{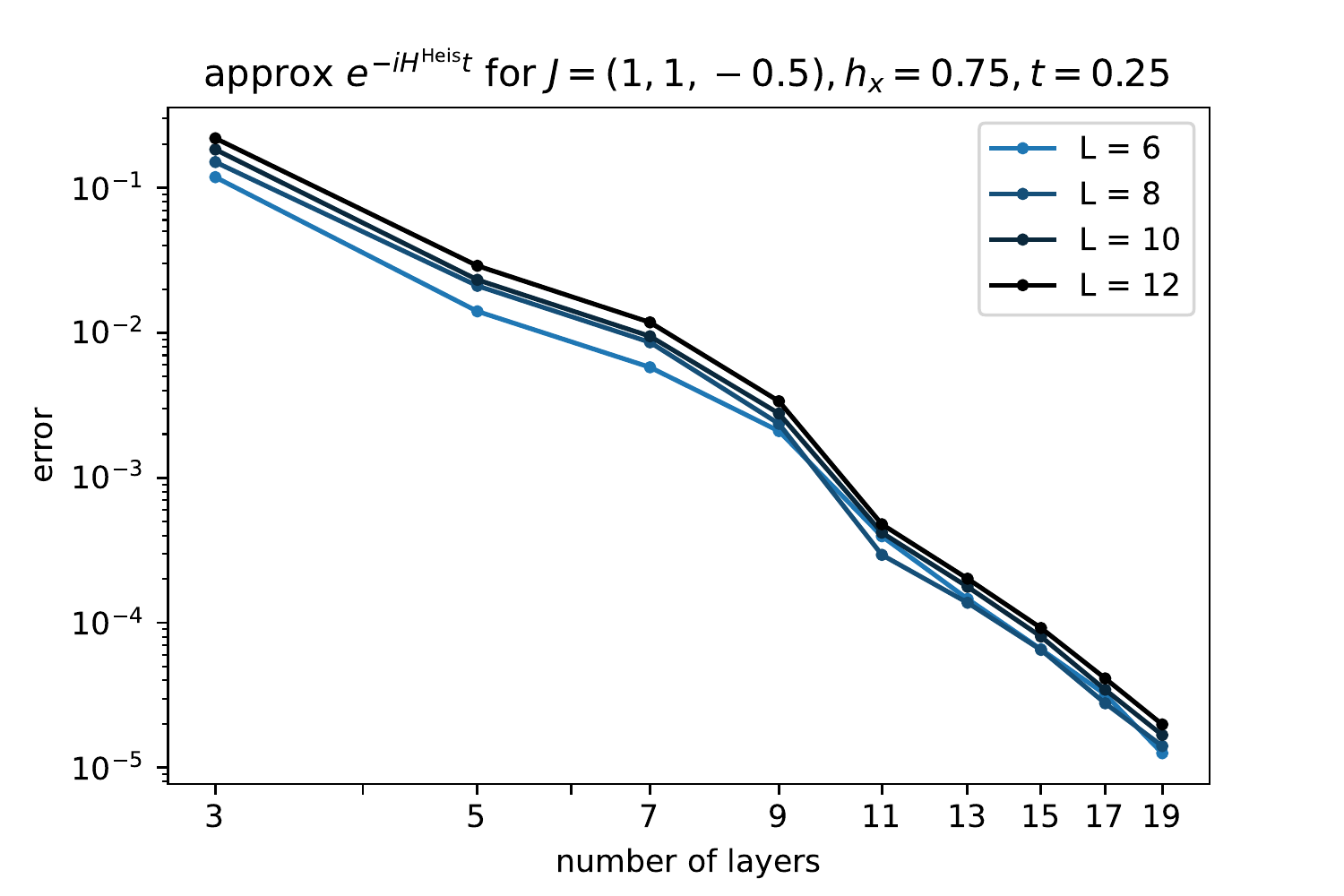}}
\caption{Approximation error of the quantum time evolution operator for $t = \frac{1}{4}$ governed by the Heisenberg model Hamiltonian \eqref{eq:H_Heisenberg} with $\vec{J} = (1, 1, -\frac{1}{2})$ and $\vec{h} = (\frac{3}{4}, 0, 0)$ on a one-dimensional lattice. (a) Comparison of the circuit optimization for $L = 6$ sites with splitting methods from the literature. (b) Approximation error for larger system sizes, using the same optimized brickwall circuit gates $(G_1, \dots, G_n)$.}
\end{figure}

The approximation error is plotted in Fig.~\ref{fig:heisenberg1d_dynamics_circuit_error}, together with a benchmark comparison of splitting methods from the literature, as before. The advantage of the optimized circuit for the Heisenberg time evolution is less pronounced than for the Ising model, but nevertheless, the error is sill more than an order of magnitude smaller as compared to the Suzuki method. The dashed curve in Fig.~\ref{fig:heisenberg1d_dynamics_circuit_error} shows the results for a more simplistic optimization protocol, namely choosing identity gates as starting points. The gap to the best results widens with increasing number of layers, which is likely due to the higher-dimensional optimization landscape. In general, one observes that the results can sensitively depend on the starting point. As described above, employing a splitting method as starting point guarantees that the optimized circuit performs at least as well as the splitting method.

As for the Ising model, we also probe the generalizability to larger systems, by showing the error depending on $L$ in Fig.~\ref{fig:heisenberg1d_dynamics_approx_larger_systems}. As expected, the error increases only slightly with $L$.

\subsection{Ising model on a ladder geometry}

As last example, we again consider the Ising model, but now on a lattice with ladder geometry, i.e., of dimension $L_x \times 2$, with periodic boundary conditions along the ladder direction. The Ising Hamiltonian (without longitudinal field) on a general lattice reads
\begin{equation}
\label{eq:H_Ising_general}
H^{\text{Ising}} = \sum_{\langle j, k \rangle} J Z_j Z_k + \sum_{j=0}^{L-1} g X_j,
\end{equation}
where the first sum runs over all nearest neighbor lattice sites. For our simulation, we set $J = 1$, $g = 3$, and final time $t = \frac{1}{4}$.

\begin{figure}[!ht]
\centering
\subfloat[comparison with splitting methods]{%
\label{fig:ising_ladder_dynamics_circuit_error}%
\includegraphics[width=\columnwidth]{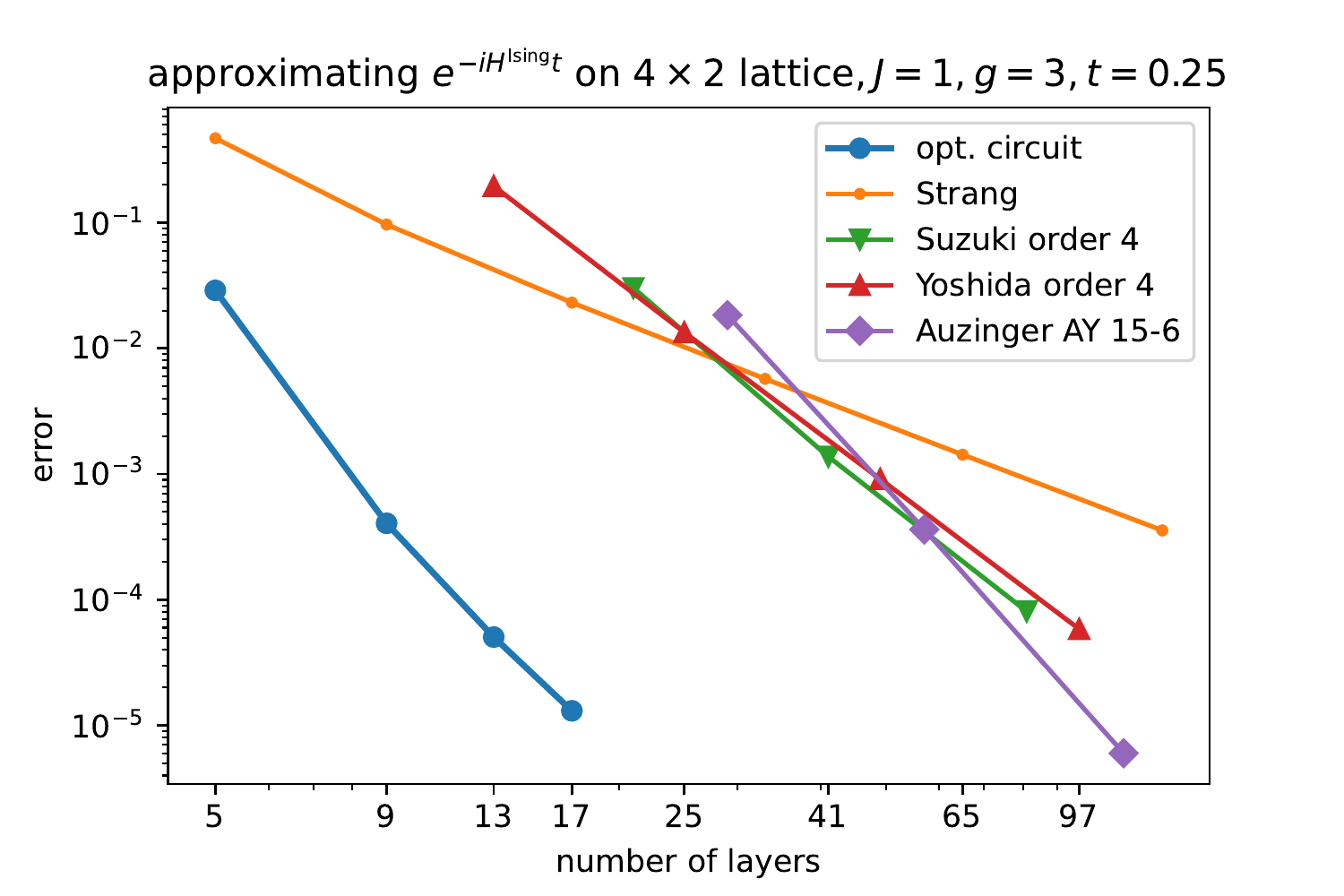}}\\
\subfloat[dependence on system size]{%
\label{fig:ising_ladder_dynamics_approx_larger_systems}%
\includegraphics[width=\columnwidth]{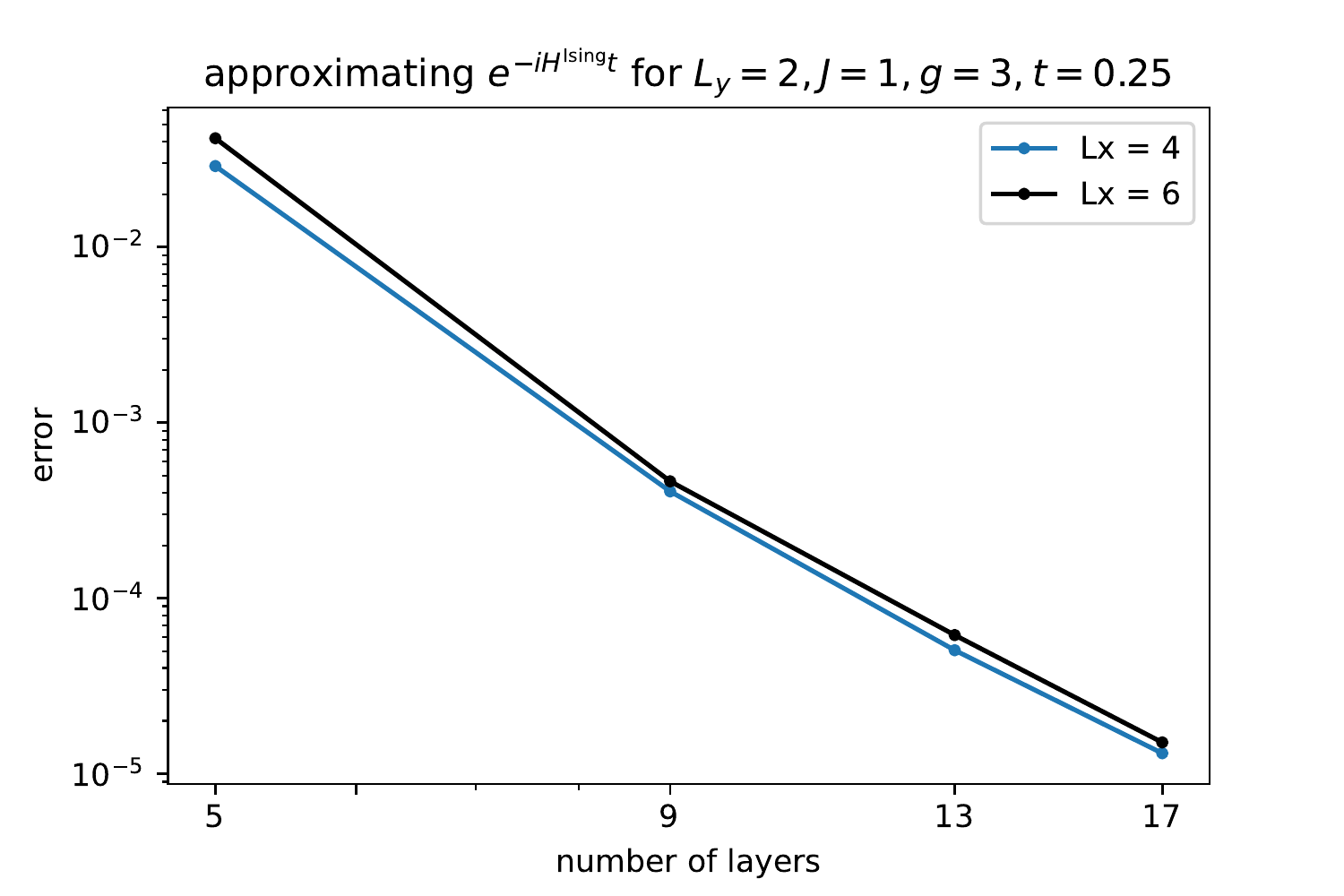}}
\caption{Approximation error of the quantum time evolution operator for $t = \frac{1}{4}$ governed by the Ising model Hamiltonian \eqref{eq:H_Ising_general} with $J = 1$ and $g = 3$ on a ladder geometry. The circuit gates have been optimized on a $4 \times 2$ lattice. (a) shows a comparison with existing splitting methods from the literature, with the blue curves corresponding to the optimized circuit of the present work. Subplot (b) displays the error for system size $6 \times 2$ as well, using the same optimized brickwall circuit gates $(G_1, \dots, G_n)$.}
\end{figure}

The ladder geometry requires a modification of the splitting scheme: besides the interactions in $x$-direction, which we can split using an even-odd scheme as before, there are now additional interactions along the steps of the ladder. Thus we split the Hamiltonian into three terms: $H^{\text{Ising}}_{\text{ladder}} = H^{\text{Ising}}_{x,\text{even}} + H^{\text{Ising}}_{x,\text{odd}} + H^{\text{Ising}}_y$. The Ansatz circuit layout for the optimization is modified accordingly, using three different gate topologies for the layers. Likewise, for the benchmark comparison we use splitting schemes supporting three terms. The Strang scheme and the methods by Suzuki and Yoshida can be adapted for this purpose. For example, the Strang splitting scheme for partitioning $H = H_{\text{a}} + H_{\text{b}} + H_{\text{c}}$ reads
\begin{multline}
\e^{-i H t} = \e^{-i H_{\text{a}} t/2} \e^{-i H_{\text{b}} t/2} \e^{-i H_{\text{c}} t} \e^{-i H_{\text{b}} t/2} \e^{-i H_{\text{a}} t/2} \\ + \mathcal{O}(t^3).
\end{multline}
In our setting, $H_{\text{a}} = H^{\text{Ising}}_{x,\text{even}}$, $H_{\text{b}} = H^{\text{Ising}}_{x,\text{odd}}$ and $H_{\text{c}} = H^{\text{Ising}}_y$. We additionally include the AY 15-6 method of order $6$ by Auzinger et al.~\cite{Auzinger2017} in the comparison. A common feature of the splitting methods and our Ansatz is the sequential ordering of the matrix exponentials and circuit layouts, respectively, namely as ``abcbabcba\dots''.

The approximation error after numerical optimization and a comparison with the existing splitting methods is shown in Fig.~\ref{fig:ising_ladder_dynamics_circuit_error}. Similar to the Ising model on a one-dimensional lattice, one observes a large advantage of the optimized circuits.

To probe the suitability of the optimized gates for larger systems, we use them for a $6 \times 2$ layout and record the approximation error in Fig.~\ref{fig:ising_ladder_dynamics_approx_larger_systems}. As expected, the error increases only slightly. We cannot reach even larger systems in this comparison due to the difficulty of computing and storing the exact matrix exponential.

\section{Conclusions and outlook}

In this work we have constructed and explored a numerical scheme for optimizing the two-qubit gates $(G_1, \dots, G_n)$ in a quantum circuit Ansatz as general unitary matrices. To use these gates in physical quantum computers still requires their further decomposition into hardware-native gates; for this purpose, one can rely on existing algorithms from the literature \cite{Bullock2003, Vatan2004}. For example, if the hardware supports CNOT and arbitrary single-qubit gates, each $G_\ell$ layer can be represented by (at most) seven layers of such gates \cite{Vatan2004}. In general, a good decomposition strategy and the ``cost'' of applying the gates will depend on the specific hardware, cf.~\cite{Clinton2021}. We leave a detailed use-case study and comparison with parametrized gates as in \cite{Mansuroglu2023a, McKeever2022} for future work.

In the numerical experiments, we have seen that the improvement due to the optimization compared to existing splitting methods strongly depends on the model. It would be interesting to gain a deeper mathematical understanding of these differences, in particular regarding the (ir-)relevance of integrability. A related question is how to find suitable starting gates for the optimization, to arrive at a global optimum in the best case. Conversely, further improvements of the optimized circuits shown in this work might be possible.

The periodic boundary conditions are a tool to avoid finite size effects in the numerical simulations. Using the optimized gates on actual quantum computers requires additional considerations regarding the boundary conditions. We note that a one-dimensional system with periodic boundary conditions (as studied in the present work) could be realized on a physical quantum computer having a two-dimensional topology and nearest-neighbor connectivity, by mapping the logical qubits to a loop formed by physical qubits. Nevertheless, endowing the logical system with open boundary conditions would require a dedicated optimization of the brick wall circuit gates without translation invariance. We leave this task for future work.

We remark that our method can also be applied to quantum models with longer-range interactions, for which an even-odd splitting of the Hamiltonian is not feasible, assuming that the conditions based on the light-cone picture in Sect.~\ref{sec:light_cone} are still satisfied. Related to that, our optimized gates could also be of practical use for the time-evolving block decimation (TEBD) algorithm, when simulating quantum systems on a classical computer.

Another natural application is the approximation of the time evolution on a two-dimensional lattice. The reasonably smallest dimension on a square lattice is a window of size $4 \times 4$, to avoid interference due to the periodic boundary conditions. Performing the numerical optimization for $16$ sites is technically challenging, but we plan to tackle this scenario via a tailored implementation, possibly running on a compute cluster. One could avoid very large matrices via a matrix-free application of circuit gates and matrix exponentials to statevectors, or via the ``exponential-free'' approach investigated in \cite{Mansuroglu2023b}.

The numerical method developed in our work can approximate a general translation invariant unitary target matrix. Thus another use-case could be the decomposition of multiple-qubit gates.

In principle, the Riemannian gate optimization framework is applicable for non-translation invariant systems as well when admitting independent gates within each layer.
Since the number of to-be optimized gates then increases with the system size $L$ (or number of ``orbitals'' in a chemistry setting), the Riemannian trust-region part of the algorithm becomes computationally more expensive in practice. More specifically, in this scenario the overall number of gates is then $n = D \cdot L/2$ for $D$ layers.

\begin{acknowledgments}
We would like to thank Alexander Kemper and Frank Pollmann for insightful discussions.

This research is part of the Munich Quantum Valley, which is supported by the Bavarian state government with funds from the Hightech Agenda Bayern Plus. The research is also supported by the Bavarian Ministry of Economic Affairs, Regional Development and Energy via the project BayQS with funds from the Hightech Agenda Bayern.
\end{acknowledgments}

\appendix

\section{Wirtinger formalism}
\label{sec:Wirtinger}

The Wirtinger derivatives are an alternative approach (as compared to Eq.~\eqref{eq:def_grad} above) to define complex derivatives. Given a complex-valued smooth function $f: \C \to \C$ (not necessarily holomorphic) and a complex number $z = x + i y$ with $x, y \in \R$, one introduces
\begin{subequations}
\begin{align}
\partial_z f(z)     &= \frac{1}{2}\left(\partial_x f - i \partial_y f \right), \\
\partial_{z^*} f(z) &= \frac{1}{2}\left(\partial_x f + i \partial_y f \right),
\end{align}
\end{subequations}
where $\partial_x f$ and $\partial_y f$ are the conventional partial derivatives when interpreting $f$ as function $f: \R^2 \to \C$, $f(x, y) = f(x + i y)$.

In case $f$ is holomorphic, the Cauchy-Riemann equations imply that the Wirtinger derivative $\partial_z f$ is equal to the complex derivative of $f$, whereas the conjugated Wirtinger derivative vanishes: $\partial_{z^*} f = 0$.

For smooth functions $f, g: \C \to \C$, the following \emph{chain rule} holds:
\begin{equation}
\label{eq:chain_rule}
\partial_z (f \circ g) = \big( \partial_w f \circ g \big) \partial_z g + \big( \partial_{w^*} f \circ g \big) \partial_z g^*,
\end{equation}
where ``$\circ$'' denotes function composition and $w = g(z)$. 

The \emph{product rule} takes the same form as for real-valued functions:
\begin{equation}
\label{eq:product_rule}
\partial_z (f \cdot g) = (\partial_z f) \cdot g + f \cdot (\partial_z g),
\end{equation}
with $f \cdot g$ the pointwise product of $f$ and $g$.

By definition, we can express the gradient in Eq.~\eqref{eq:def_grad} of a smooth function $f: \C \to \R$ in terms of the Wirtinger derivative:
\begin{equation}
\label{eq:gradient_Wirtinger_relation}
\grad f(z) = 2 \big(\partial_z f(z) \big)^*.
\end{equation}

\bibliography{references}

\end{document}